\documentclass[review,authoryear,12pt]{elsarticle}

\usepackage{mathtools,amsmath,amssymb}
\usepackage{mathrsfs,bbm}
\usepackage{empheq}
\usepackage[a4paper,text={16.5cm,25.2cm},centering]{geometry}

\usepackage{graphicx}

\usepackage{subfig}
\usepackage[colorlinks,linkcolor=black,citecolor=black,urlcolor=black]{hyperref}
\usepackage{booktabs}
\usepackage{multirow}

\newtheorem{definition}{Definition}

\newtheorem{lemma}{Lemma}

\newtheorem{theorem}{Theorem}
\newtheorem{remark}{Remark}

\newcommand{\Exp}{\mathbb{E}}

\newcommand{\Cov}{\mathbb{C}\text{ov}}

\newcommand{\Var}{\mathbb{V}\text{ar}}

\newcommand{\norm}[1]{\left\lVert #1 \right\rVert}

\newcommand{\map}[2]{\,{:}\,#1\!\longrightarrow\!#2}

\newcommand{\innerp}[2]{ \langle{#1} , {#2}\rangle }

\newcommand{\Besov}{\mathcal B}

\def\K{K}

 \usepackage{lineno}

\journal{}

\begin{document}

\begin{frontmatter}



\title{Wavelet Deconvolution in a Periodic Setting with Long-Range Dependent Errors}


\author{Justin Rory Wishart\fnref{fn1}}
\ead{j.wishart@unsw.edu.au}
\address{Department of Mathematics \& Statistics \\
The University of Melbourne \\
Parkville, VIC, 3010 }
\fntext[fn1]{The author has moved since the work was completed. The present address of the author is at the University of New South Wales.}

\begin{abstract}
In this paper, a hard thresholding wavelet estimator is constructed for a deconvolution model in a periodic setting that has long-range dependent noise. The estimation paradigm is based on a maxiset method that attains a near optimal rate of convergence for a variety of $\mathscr L_p$ loss functions and a wide variety of Besov spaces in the presence of strong dependence. The effect of long-range dependence is detrimental to the rate of convergence. The method is implemented using a modification of the {\tt WaveD}-package in {\tt R} and an extensive numerical study is conducted. The numerical study supplements the theoretical results and compares the LRD estimator with na\"ively using the standard {\tt WaveD} approach.
\end{abstract}


\end{frontmatter}
\graphicspath{{./Plots}}
 \linenumbers

\section{Introduction}
Nonparametric estimation of a function in a deconvolution model has been studied widely in various contexts. We study the deconvolution model with a long range dependent (LRD) error structure. More specifically, we consider the problem of estimating a function $f$ after observing the process,
\begin{equation}
	dY(x) = \K * f(x)\, dx + \varepsilon^\alpha dB_H(x), \qquad x \in [0,1];
\label{eq:fWnmodel}
\end{equation}
where $\K * f(x) = \int_0^1 K(t) f(x-t) \, dt$ is the regular convolution operator, $\varepsilon \asymp n^{-1/2}$, $B_H$ is a fractional Brownian motion. The fractional Brownian motion is defined as a Gaussian process with zero mean and covariance structure,
\[
	\Exp B_H(t) B_H(s) = \frac{1}{2} \left( |s |^{2H} + |t|^{2H} - |t - s|^{2H}\right)
 \] and $\alpha = 2 - 2H \in (0,1]$ is the level of long-range dependence (where $H$ denotes the standard Hurst parameter). The assumption of an i.i.d. error structure is captured as a special case of \eqref{eq:fWnmodel} with the choice $\alpha = 1$ which reduces the model to a standard Brownian motion structure. The convolution operator $\K$ is assumed to be of the {\it regular-smooth} type such that the Fourier coefficients
\begin{equation}
	 |\widetilde \K[\omega]| \asymp |\omega|^{-\nu}\qquad \text{for all }\omega \in \mathbb{R},\label{eq:regular}
\end{equation}
where $\nu \ge 0$ and $\widetilde \K[\omega]$ denotes the Fourier transform $\widetilde \K[\omega] = \mathcal F \K [\omega] \coloneqq \int_\mathbb{R} e^{-2\pi i \omega x} \K(x)\, dx.$ 

Deconvolution is a common problem occuring in several areas such as econometrics, biometrics, medical statistics and image reconstruction. For example, the method can be applied to the light detection and ranging (LIDAR) techniques and image de-blurring techniques. The parameter $\nu > 0$ is often referred to as the {\it degree of ill-posedness} (DIP) with $\nu = 0$ denoting the {\it well-posed} or direct case.  

Various wavelet methods have been constructed to address the deconvolution problem over the last two decades (see for example \cite{Donoho-1995,Wang-1997,Abramovich-Silverman-1998,Walter-Shen-1999,Fan-Koo-2002, Donoho-Raimondo-2004,Johnstone-Raimondo-2004,Johnstone-et-al-2004,Kalifa-Mallat-2003,Pensky-Sapatinas-2009}). 

In the standard deconvolution models, the assumption of i.i.d. variables is made. However, empirical evidence has shown that even at large lags, the correlation structure in variables can decay at a hyperbolic rate. To account for this, an extensive literature on LRD variables has emerged to describe this phenomena. Areas of applications of LRD analysis include economics with financial returns, volatility and stock trading volumes; hydrology in rainfall and temperature data; and computer science with data network traffic data. There are many more applications of LRD analysis and the interested reader is referred to \cite{Beran-1992,Beran-1994} and \cite{Doukhan-et-al-2003} for more details. Some analysis has been done for the direct model ($\nu = 0$) with LRD errors in works such as \cite{Wang-1996,Kulik-Raimondo-2009}. The topic of density deconvolution with LRD has been studied by \cite{Kulik-2008,Chesneau-2012}.

The aim of this paper is to study a wavelet deconvolution algorithm that can be easily applied in the context of a deconvolution problem with LRD errors as in model \eqref{eq:fWnmodel}. Minimax rates of estimation of $f$ have been established in our context by \cite{Wang-1997} for the squared-error loss. However, their method uses the Wavelet-Vaguelette Decomposition (WVD) which is a sophisticated transform where, to the authors knowledge, there is no freely available software for implementation.

There are two main contributions that this paper will address. The first contribution will establish theoretical results for a wide variety of function classes over many error measures (which includes the squared-error loss considered in \cite{Wang-1997} as a special case) by adapting the approaches of \cite{Johnstone-et-al-2004} and \cite{Kulik-Raimondo-2009a}. The estimation of $f$ can be achieved with an accuracy of order,
\[
	\left( \frac{\log n}{n} \right)^{\rho}, \qquad \rho = \frac{\alpha s p}{2s + 2\nu + \alpha},
\]
where the performance is measured by the error loss from the $\mathscr L_p$ metric. The approach of \cite{Johnstone-et-al-2004} used a hard thresholding wavelet estimator. We modify their approach by determining the appropriate threshold levels and fine scale level under the strong dependence structure in \eqref{eq:fWnmodel}. 

\begin{figure}[!ht]
\centering
\begin{minipage}{\columnwidth}
\subfloat[LIDAR signal.]{
\label{comparisonfiga}
\includegraphics{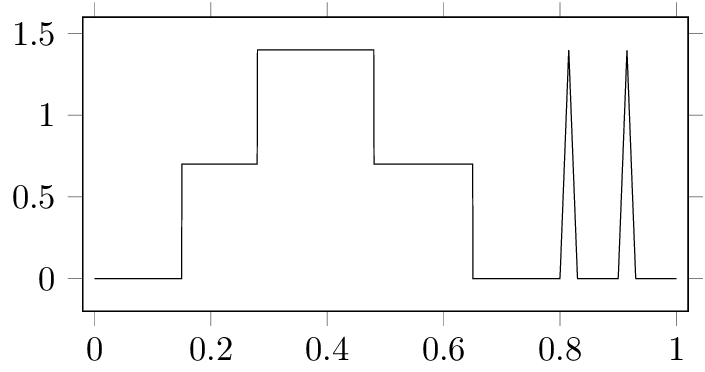}
}
\subfloat[Doppler signal.]{
\label{comparisonfigb}
\includegraphics{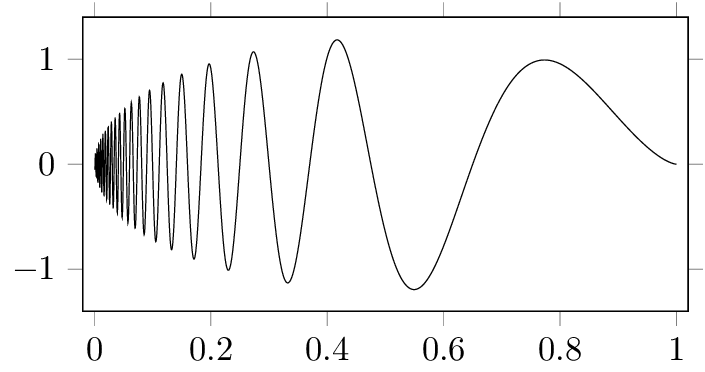}
}
\end{minipage}
\begin{minipage}{\columnwidth}
\subfloat[Cusp signal.]{
\label{comparisonfigc}
\includegraphics{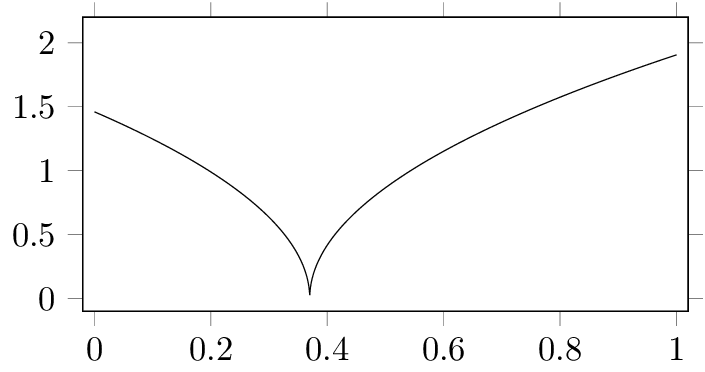}
}
\subfloat[Bumps signal.]{
\label{comparisonfigd}
\includegraphics{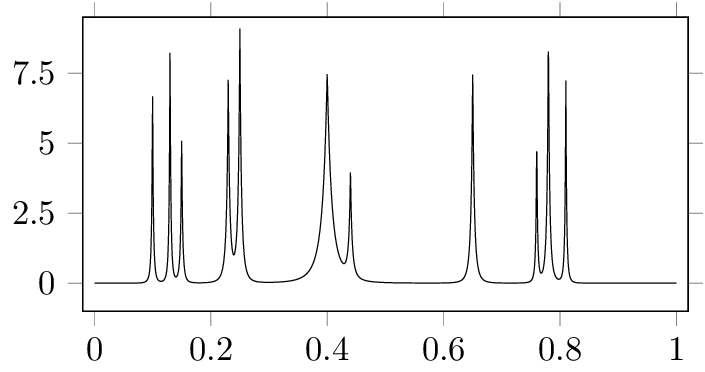}
}
\end{minipage}
\caption{Four signals that are common in signal processing and deconvolution.}
\label{fig:signals}
\end{figure}

The second contribution is allowing an easily implementable method for estimation in practice by modifying the already established {\tt WaveD} approach of \cite{Raimondo-Stewart-2007}. The {\tt WaveD} software is freely available from CRAN (\url{http://cran.r-project.org/}). With our modification of {\tt WaveD}, a numerical study is conducted comparing the performance of the default {\tt WaveD} method and the LRD method presented here. Four popular test case signals are used to benchmark methods in the literature are the Doppler, LIDAR, Bumps and Cusp signals. These are used here and are shown in \autoref{fig:signals}.

\begin{figure}[!ht]
\centering
\begin{minipage}{\columnwidth}
\subfloat[LIDAR signal.]{
\label{comparisonfiga}
\includegraphics{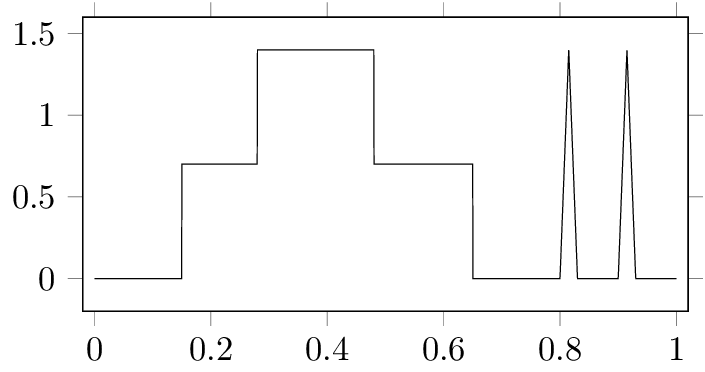}
}
\subfloat[Cusp signal.]{
\label{comparisonfigb}
\includegraphics{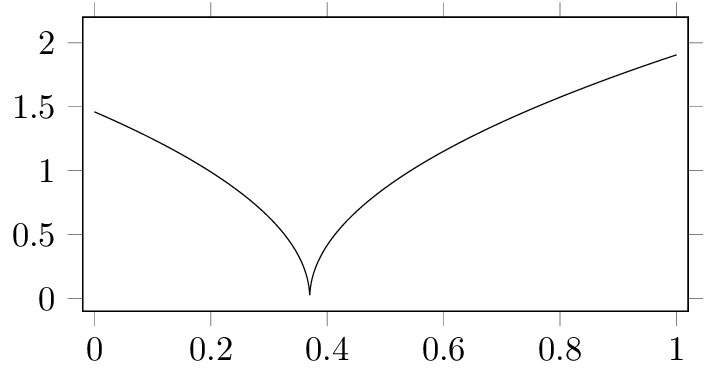}
}
\end{minipage}
\begin{minipage}{\columnwidth}
\subfloat[Blurred signal with $\nu = 0.7,\, \alpha = 0.5$.]{
\label{comparisonfigc}
\includegraphics{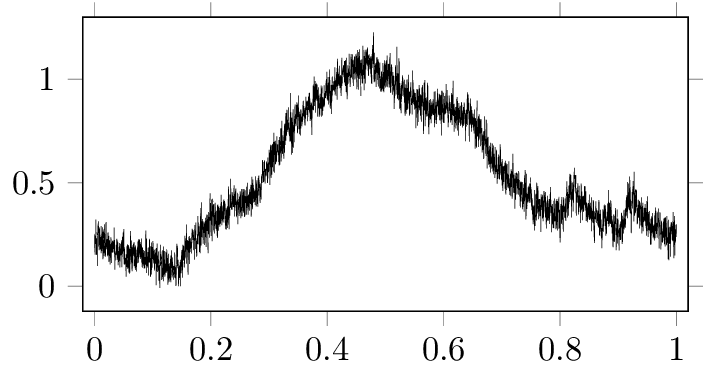}
}
\subfloat[Blurred signal with $\nu = 0.7,\, \alpha = 0.6$.]{
\label{comparisonfigc}
\includegraphics{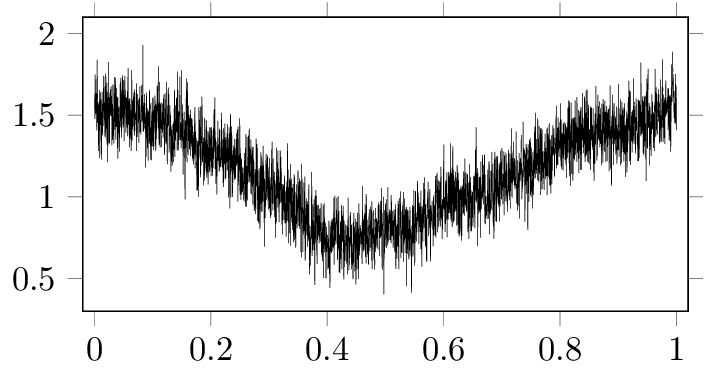}
}
\end{minipage}
\begin{minipage}{\columnwidth}
\subfloat[Default {\tt WaveD} reconstruction, $\widehat j_{1,1} = 6$.]{
\label{comparisonfigc}
\includegraphics{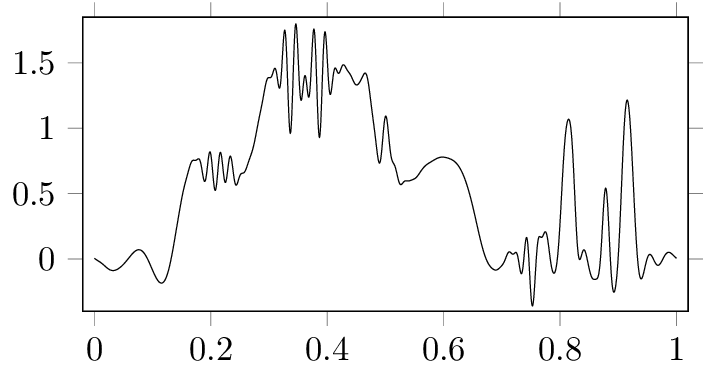}
}
\subfloat[Default {\tt WaveD} reconstruction, $\widehat j_{1,1} = 6$.]{
\label{comparisonfigc}
\includegraphics{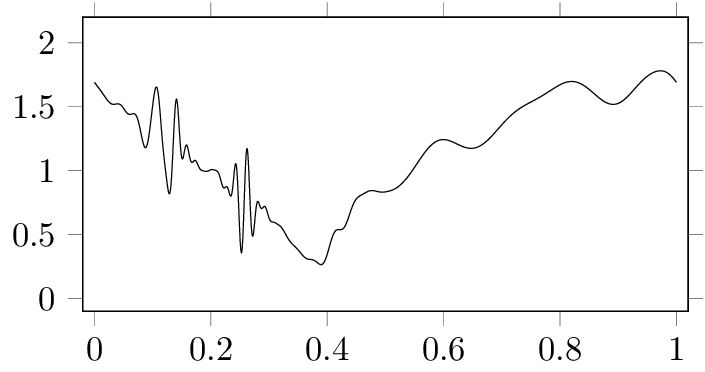}
}
\end{minipage}
\begin{minipage}{\columnwidth}
\subfloat[LRD {\tt WaveD} reconstruction, $\widehat j_{\alpha,1} = 5$.]{
\label{comparisonfigc}
\includegraphics{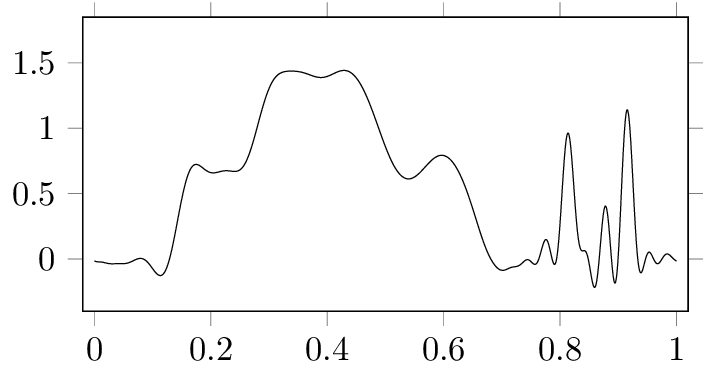}
}
\subfloat[LRD {\tt WaveD} reconstruction, $\widehat j_{\alpha,1} = 4$.]{
\label{comparisonfigc}
\includegraphics{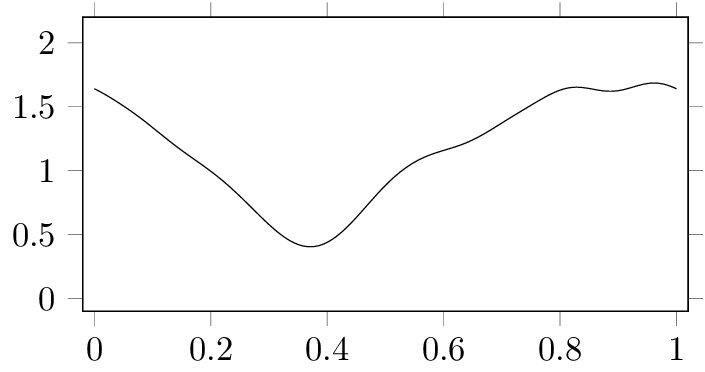}
}
\end{minipage}
\caption{Comparison of the LRD {\tt WaveD} method against the default {\tt WaveD} method. In both cases, $n = 4096 = 2^{12}$. The LRD method truncates the finest scale earlier as noted. For the LIDAR and cusp signals, SNR $\approx$ 20dB. }
\label{fig:comparisons}
\end{figure}

As will become evident in the later in the theoretical analysis. The case of LRD errors introduces some challenges to the estimation. For strong dependence there can be artificial trends in the noise which require modified thresholds in the wavelet estimator compared to the i.i.d. case. These effects are demonstrated clearly in \autoref{fig:comparisons} for the cusp signal whereby the addition of a higher scale in the wavelet estimation deteriorates the performance of the standard {\tt WaveD} method when there is a strong LRD error structure. Since the finest scale is too large, the i.i.d. method also includes the spurious trends which were artificially generated by the LRD error sequence.
\subsection{Outline}
\label{ssec:outline}

A review of periodised Meyer wavelets and the Besov functional class are given in \autoref{sec:preliminary}. In \autoref{sec:estimationmethod} the basic argument for the deconvolution technique is given along with the convergence rate results. In \autoref{sec:Sim}, the method is implemented in {\tt R} and a numerical study is conducted to confirm the rate results and a compare with the current {\tt WaveD} methodology. The mathematical proofs are given in \autoref{sec:proof}.

\section{Preliminary framework}
\label{sec:preliminary}

\subsection{Periodised Meyer wavelet basis}
\label{ssec:Meyer}
Let $\left( \phi, \psi \right)$ denote the Meyer wavelet scaling and detail basis functions defined on the real line $\mathbb{R}$; (see \cite{Meyer-1992} and \cite{Mallat-1999}). These are defined in the Fourier domain with, 
\[
\widetilde{\phi}(\omega) = \begin{cases}
1, & \text{if }|\omega| \le 1/3\\
\cos (\pi/2 v(3 |\omega| - 1)), & \text{if } |\omega| \in (1/3,2/3]\\
0, &\text{otherwise}.
\end{cases}
\]
and the mother wavelet function defined with, 
\begin{equation} \widetilde{\psi}(\omega) = \begin{cases}
e^{ - i \pi \omega }\sin (\frac{\pi}{2} v(3 |\omega| - 1)), & \text{if }|\omega| \in (1/3,2/3]\\
e^{ - i \pi \omega }\cos (\frac{\pi}{2} v(3/2 |\omega| - 1)), & \text{if } |\omega| \in (2/3,4/3]\\
0, &\text{otherwise}.
\end{cases}\label{eq:MeyerBandlimited}\end{equation}
The auxiliary function $v$ is a piecewise polynomial that can be chosen to ensure that the Meyer wavelet has enough vanishing moments. For the bivariate index $(j,k)$, the dilated and translated mother and father wavelets at resolution level $j$ and time position $k$ are defined,
\[
   \phi_{j,k}(x) = 2^{j/2}\phi(2^jx-k), \quad  \psi_{j,k}(x) = 2^{j/2}\psi(2^jx-k), \quad \text{for }j \ge 0.
\]
For our purposes, we are interested with a multiresolution analysis for periodic functions on $\mathscr L_2([0,1])$. This is done by periodising the Meyer basis functions with,
\[  \Phi_{j,k} (x) = \sum_{l \in \mathbb{Z}} \phi_{j,k}(x + l) \quad   \text{and} \quad \Psi_{j,k} (x) = \sum_{l \in \mathbb{Z}} \psi_{j,k}(x + l). \] 
Consequently, any $1-$periodic function $f \in \mathscr L_2([0,1])$ can be written with a wavlet expansion with,
\[
  f(x) = \sum_{k = j_0}^{2^{j_0}-1} \alpha_{j_0,k} \Phi_{j_0,k}(x) + \sum_{j=j_0}^\infty \sum_{k = 0}^{2^j-1} \beta_{j,k} \Psi_{j,k}(x),
\]
where $\alpha_{j_0,k} = \innerp{f}{\Phi_{j,k}} = \int_0^1 f(t) \Phi_{j_0,k}(t) \, dt$ and $\beta_{j,k} = \innerp{f}{\Psi_{j,k}} = \int_0^1 f(t) \Psi_{j,k}(t) \, dt$.

This particular expansion is used since the Meyer wavelet is bandlimited (see \eqref{eq:MeyerBandlimited}) which allows the use of efficient algorithms with \cite{Raimondo-Stewart-2007} and for mathematical convenience in the later proofs.

\subsection{Functional Class}
\label{sec:functionalclass}

We analyse the estimation procedure over a Besov space of periodic functions which have a nice characterisation using the coefficients of a wavelet expansion (granted that the wavelet is periodic and has enough smoothness and vanishing moments).
\begin{definition}
	\label{def:Besov}
	For $f \in \mathscr L_\pi([0,1])$ with $ \pi \ge 1$ and $r \ge 1$,
\[
	f = \sum_{j,k} \beta_{j,k}\Psi_{j,k}(x) \in \Besov_{\pi,r}^s \Leftrightarrow \sum_{j \ge 0} 2^{j(s + \tfrac{1}{2} - \tfrac{1}{\pi})r}\left( \sum_{k = 0}^{2^j} \left| \beta_{j,k}\right|^\pi \right)^{\tfrac{r}{\pi}} < \infty.
\]
\end{definition}
The consideration of a Besov class allows a more precise analysis of the asymptotic convergence results of the signal $f$ since the Besov class includes other common functional classes as a special case. Loosely speaking, a function $f \in \Besov_{\pi,r}^s$ includes functions that are $s$ times differentiable with $f^{(s)} \in \mathscr L_\pi(0,1)$. The parameter $r$ is less important, it allows a more intricate variation of behaviour in the Besov class. The special cases when $\pi$ or $r = \infty$ involve replacing the $\ell_\pi$ and $\ell_r$ type norms in \autoref{def:Besov} with the $\ell_\infty$ norm. For example, if $f \in \Besov^s_{\infty,\infty} \Rightarrow \sup_{j \ge 0} 2^{j(s+1/2-1/\pi)} \sup_{(j,k)}|\beta_{j,k}| < \infty.$ The interested reader is referred to \cite{Donoho-et-al-1995,Donoho-Johnstone-1998} and references therein for a more detailed discussion.
\section{Function estimation method}
\label{sec:estimationmethod}

We use the wavelet shrinkage paradigm for our estimation which has become a standard statistical procedure in nonparametric estimation. We use a hard thresholding approach where the wavelet estimator is defined,
\begin{equation}
	\widehat f (x) =\sum_{k = 0}^{2^{j_0}-1} \widehat \alpha_{j_0,k} \mathbbm{1}_{\left\{ |\widehat    \alpha_{j_0,k} | \ge \lambda_{j_0} \right\}}\Phi_{j_0,k}(x) + \sum_{(j,k) \in \Lambda_n}^{2^j-1} \widehat \beta_{j,k} \Psi_{j,k}(x), \label{eq:waveletestimator}
\end{equation}
where $\widehat \alpha_{j_0,k}$ and $\widehat \beta_{j_0,k}$ are the estimated wavelet coefficients. The level $j_0$ corresponds to the coarse resolution level and the set of indices $\Lambda_n = \left\{(j,k): j_0 \le j \le j_1; k = 0,1,\ldots, 2^j-1 \right\}$ indexes details of the function up to a fine resolution level $j_1$. The wavelet estimation procedure keeps only the coefficients in the expansion when $\left| \widehat \beta_{j,k} \right| \ge \lambda_j$, a scale dependent threshold. The parameters $\lambda_j$ and indices $\Lambda_n$ are chosen to control the noise embedded in the estimated wavelet coefficients in an optimal way in the sense of the rate results presented in the next section.

For the deconvolution problem it is natural to conduct analysis in the Fourier domain since the deconvolution operator becomes a multiplier in the Fourier domain. The deconvolution model, \eqref{eq:fWnmodel}, has Fourier domain representation with,
\[
	\widetilde{Y}[n] \coloneqq \int_\mathbb{R} e^{-2\pi i n x}\, dY(x) = \widetilde{\K}[n] \widetilde f[n] + \varepsilon^\alpha \widetilde Z_H[n]. \label{eq:FourierfWnmodel}
\]
Then using the Parseval identity, one can obtain a representation of the wavelet coefficients.
\begin{align}
	\widehat{\beta}_{j,k} &= \sum_{n \in \mathbb{Z}} \frac{\widetilde{Y}[n]}{\widetilde \K[n]} \overline{\widetilde{\Psi_{j,k} }}[n]\nonumber\\
		&=  \sum_{n \in \mathbb{Z}} \widetilde f[n] \overline{\widetilde{\Psi_{j,k} }}[n] + \varepsilon^\alpha \sum_{n \in \mathbb{Z}} \frac{\widetilde{Z_H}[n]}{\widetilde \K[n]} \overline{\widetilde{\Psi_{j,k} }}[n]\nonumber\\
		&= \beta_{j,k} + \varepsilon^\alpha \sum_{n \in \mathbb{Z}} \frac{\widetilde{Z_H}[n]}{\widetilde \K[n]} \overline{\widetilde{\Psi_{j,k} }}[n]
		\label{eq:beta}
\end{align}
The Meyer wavelet is bandlimited (see \eqref{eq:MeyerBandlimited}) meaning that the sums are finite. In particular, define the summation sets for the detail coefficients at scale level $j$ with,
\begin{equation}
  \mathbb{ D}_j \coloneqq \left\{ z = \pm a :  a\in \left\{ \lceil 2^j/3 \rceil, \lceil 2^j/3 \rceil + 1, \ldots, \lfloor 2^{j+2}/3 \rfloor \right\} \right\},\label{eq:Meyerdomain}
\end{equation}
which has cardinality $|\mathbb{D}_j|  = 2^{j+1}$. A similar procedure is conducted to estimate the scale coefficients $\alpha_{j_0,k}$ by using the Fourier coefficients of the scale function $\Phi_{j_0,k}$ instead of the detail function $\Psi_{j,k}$.

\subsection{Maxiset Approach}
\label{Maxiapproach}

The methodology used here is an amalgamation of the deconvolution work of \cite{Johnstone-et-al-2004} and the LRD work of \cite{Kulik-Raimondo-2009}. The nonlinear wavelet estimator \eqref{eq:waveletestimator} can be analysed using the maxiset approach with the level dependent threshold $\lambda = \lambda_j$ and fine resolution level $j_1$ which depends on $\alpha$ and $\nu$.

\textit{Fine resolution level}. The range of resolution levels (frequencies) where the estimator \eqref{eq:waveletestimator} is considered is,
\begin{equation}
	\Lambda_n = \left\{ (j,k) : j_0 \le j \le j_1, 0 \le k \le 2^j\right\}.\label{eq:range}
\end{equation}
The finest resolution level $j_1$ is set to be,
\begin{equation}
	2^{j_1} = 2^{j_{\alpha,1}} \coloneqq \left( \frac{n^\alpha}{\log n}\right)^{1/(\alpha+2\nu)},\label{eq:resolution}
\end{equation}
Then consider the precise form of the level dependent thresholds in $\lambda = \lambda_j$. As in previous works, this level dependent threshold will have three input parameters, written,
\begin{equation}
	\lambda_j = \xi \sigma_{j,\nu,\alpha} c_n,
	\label{eq:threshold}
\end{equation}
where the three values are given by,
\begin{itemize}
	\item $\xi$ : a constant that depends on the tail of the noise distribution. Theoretically this should satisfy the bound $\xi > 2\sqrt{ \alpha (p \vee 2)}$.
	\item $\sigma_j$: the level-dependent scaling factor that is based on the convolution kernel and level of dependence,
	\begin{equation}
		\sigma_j = \sigma_{j,\nu,\alpha} = \mathcal O(2^{-j(1-\alpha-2\nu)/2}).
		\label{eq:scalefactor}
	\end{equation}
	\item $c_n$ : a sample size dependent scaling factor,
	\begin{equation}
		c_n = \sqrt{n^{-\alpha}\log n}.
		\label{eq:samplefactor}
	\end{equation}
\end{itemize}
The smoothing parameter is intentionally denoted $\xi$ to distinguish it clearly from the smoothing parameter, denoted $\eta$ used in the i.i.d. case in \cite{Johnstone-et-al-2004} and its {\tt WaveD} implementation in \cite{Raimondo-Stewart-2007}.

\begin{theorem}
	\label{Theorem}
	Consider model \eqref{eq:fWnmodel} with the wavelet estimator \eqref{eq:waveletestimator} with the coefficient estimates in \eqref{eq:beta} using the thresholds and resolution levels given by, \eqref{eq:range}, \eqref{eq:resolution}, \eqref{eq:threshold}, \eqref{eq:scalefactor} and \eqref{eq:samplefactor}. Assume that $p > 1$ and $f \in \mathcal B^s_{\pi,r}$ with $\pi \ge 1$, $s \ge 1/\pi$ where $r$ satisfies
 \[
	0 < r \le \min \left\{ \frac{(2\nu + \alpha)p}{2s + 2 \nu + \alpha}, \frac{(2\nu + \alpha)p - 2}{2s + 2 \nu - \frac{2}{\pi}\alpha}\right\}.
\]
Then there exists a constant $C>0$ such that,
\[
	\Exp \norm{f - \widehat f}_p^p \le C \left( \frac{\log n}{n}\right)^{ \rho  },
\]
where $\norm{\cdot}_p$ is the standard $p$-norm and,
\begin{empheq}[left={\rho=\empheqlbrace}]{alignat=2}
	& \frac{ \alpha s p }{2s + 2\nu + \alpha}, && \quad\text{if }\quad s \ge (2 \nu + \alpha)\left( \frac{p}{2\pi} - \frac{1}{2}\right);\label{eq:dense}\\
	& \frac{ \alpha p(s-\tfrac{1}{\pi} + \tfrac{1}{p}) }{2s + 2\nu + \alpha- \frac{2}{\pi}}, && \quad \text{if }\quad\frac{1}{\pi}- \nu - \frac{\alpha}{2} < s < (2 \nu + \alpha)\left( \frac{p}{2\pi} - \frac{1}{2}\right).\label{eq:sparse}
\end{empheq}
\end{theorem}

\begin{remark}
\label{rem:sparse}
There is an `elbow effect' or `phase transition' in the rates of convergence switching from \eqref{eq:dense} to \eqref{eq:sparse} which are usually referred to as the `dense' and `sparse' phases respectively. This is similar to the effect seen in \cite{Kulik-Raimondo-2009a} and \cite{Johnstone-et-al-2004}. The fBm errors has increased the size of the dense region in comparison to the standard Brownian motion when the boundary was at $s = (2\nu + 1)(p/2\pi - 1/2)$ (see \cite{Johnstone-et-al-2004}). Also for the sparse region, the condition $p > 2/(2 \nu + \alpha)$ is needed to ensure that it is well defined. When $2\nu + \alpha \ge 2$, this is not a restriction since it is assumed that $p > 1$.
\end{remark}

\begin{remark}
The rate results are consistent with works on LRD and inverse problems in the literature in the sense that the rate of convergence deteriorates as $\alpha$ decreases (stronger dependence) or the DIP parameter $\nu$ increases (more ill-posed). In particular, our rate results agree with the results obtained in \cite{Johnstone-et-al-2004} in the i.i.d. deconvolution setting with smooth convolution, $|\widetilde \K(\omega)| \sim |\omega|^{-\nu}$, with the choice $\alpha = 1$. Our rate results also agree with the results obtained in \cite{Kulik-Raimondo-2009a} with the direct regression model with LRD errors with the choice $\nu = 0$. The results agree with minimax results the for the squared-error loss ($p=2$) by \cite{Wang-1997}.
\end{remark}

\begin{remark}
Our estimator is adaptive with respect to the smoothness parameter $s$ since the tuning paradigm does not depend on $s$. However it is not adaptive with respect to the LRD parameter $\alpha$ since the level dependent thresholds, \eqref{eq:resolution}, \eqref{eq:scalefactor} and \eqref{eq:samplefactor} depend on $\alpha$.
\end{remark}

\section{Numerical Study}
\label{sec:Sim}

A numerical study is now considered with the focus being the effect of the dependence structure. Ideally, it would be desirable to compare the LRD {\tt WaveD} method introduced here with the minimax optimal WVD method considered by \cite{Wang-1997}. However, to the authors knowledge, no freely available implementation of the WVD method exists.

One of the immediate challenges of implementing the LRD approach is that $\alpha$ is unknown in practice. The estimation of $\alpha$ is challenging problem in its own right that has been studied in the literature, see for example \cite{Veitch-Abry-1999} and more recently \cite{Park-Park-2009} in this regard. Having full data driven estimates of all the parameters is beyond the scope of this work and we will assume that $\alpha$ is known. For our purposes we wish to compare the performance of the regular {\tt  WaveD} approach which was designed for i.i.d. observations with the LRD extension suggested here.

The default {\tt WaveD} method uses a stopping rule in the Fourier domain using the results of \cite{Cavalier-Raimondo-2007} to estimate the finest permissible scale level in the expansion. This concerns the case of model \eqref{eq:fWnmodel} with a standard Brownian motion ($\alpha = 1$) where $ 2^{j_1} = 2^{j_{1,1}}= (n/\log n)^{1/(1+ 2\nu)}$. For a fair comparison, the finest permissible scale level, $j_{\alpha,1}$, should be estimated in the same vein for our LRD extension. The Fourier stopping rule is extended to the LRD framework below. 

Briefly reviewing the fine scale estimation method of \cite{Cavalier-Raimondo-2007}, the scenario of an unknown convolution kernel $\K$ is considered. To alleviate this they assume that it is possible to choose input signals $f$ in \eqref{eq:fWnmodel} to gain information about $\K$. In particular, the Fourier basis is chosen $f = \left( e_\ell \right)_{\ell \in \mathbb{Z}}$ and passed into the deconvolution model. Doing so here, one would pass the Fourier basis into \eqref{eq:fWnmodel} and denote this new information with $dY_e(x) = \K * e_{\ell} (x) + \varepsilon^\alpha dB_H(x)$. Due to the orthogonality of the Fourier basis, the Fourier domain representation of $Y_e$ is
\[
	\widetilde Y_e[\ell] = \widetilde \K [\ell] + \varepsilon^\alpha W_H[\ell]
\]
where $\widetilde W_H[\ell]$ is a complex Gaussian random variable identically distributed to $\widetilde Z_H[\ell]$ but independent of $\widetilde Z_H[\ell]$. We then estimate the fine scale level by, 
\begin{equation}
	\widehat j_{\alpha,1} = \lfloor \log_2 M \rfloor -1
	\label{eq:finescaleestimator}
\end{equation}
where the stopping time $M$ is determined in the Fourier domain with,
\[
	M = \min\left\{ \ell, \ell > 0 : |\widetilde Y_e[\ell]| \le \ell^{\alpha/2} \varepsilon^{\alpha} \log 1/\varepsilon^{2} \right\},
\]
and $\lfloor x \rfloor $ is the largest integer small than $x$. The estimate $\widehat j_{\alpha,1}$ is close to $j_{\alpha,1}$ with high probability due to \autoref{lem:highlevel} in \autoref{sec:proof}.

\subsection{Implementation procedure}
\label{ssec:implement}

We are now in a position to be able to implement an estimation procedure for the LRD model \eqref{eq:fWnmodel} using a modification of {\tt WaveD} method of \cite{Raimondo-Stewart-2007}. This is conducted using the test functions shown in \autoref{fig:signals}. A discretely sampled deconvolution model is repeatedly simulated with these test signals and the {\tt WaveD} and LRD modification estimates computed. The performance measure is the Mean-Square Error (MSE) which is calculated empirically over $M = 1024$ repeated simulations.
\begin{enumerate}
	\item Choose $f(t)$ to be a LIDAR, Doppler, Bumps or Cusp signal (see \autoref{fig:signals}) over the grid $t_i = i/n$ with $i = 1,2,\ldots, n$ and $n = 2^{12} = 4096$. The LIDAR and Doppler signals were generated from by the code from the {\tt WaveD} package and the Bumps and Cusp signal code was imported into {\tt R} from the {\tt WaveLab} package in {\tt MATLAB}. The signals were standardised to agree with the signal levels in \cite{Cavalier-Raimondo-2007}.
	\item Simulate the LRD error process using the {\tt fracdiff} package of {\tt R} available from {\tt CRAN}. Use the {\tt fracdiff.sim} command to simulate a FARIMA sequence, the  parameters kept at their defaults with the dependence controlled with $d = (1-\alpha)/2$ and the LRD error process was standardised to have unit variance.
	\item Generate the convolution kernel, $\K = k$, to be the Gamma density with scale parameter $0.25$ and shape parameter $0.7$ (The $DIP$ in this case is $\nu =0.7$).
	\item Generate the data $y_i = k * f (t_i) + \sigma 2^{-\alpha/2}e_i$ where $e_i$ is the FARIMA sequence. The size of $\sigma$ is governed by the blurred signal-to-noise ratio (SNR) where 
	\[
		\text{SNR} = 10 \log_{10} \left( \norm{\K * f}^2/\sigma^2 \right)
	\]
	The SNR is considered for three scenarios SNR = 10dB (high noise), 20dB (medium noise) or 30dB (low noise).
	\item Compute the default {\tt WaveD} estimator which assumes three vanishing moments and starts the wavelet expansion at scale level $j_0 = 3$. The level dependent thresholds are computed, $\lambda_j = \eta \tau_j c_n$ where $\eta = \sqrt{6}$, $c_n = \widehat \sigma \sqrt{\log n/n}$ and $\tau_j = (|\mathbb{D}_j|^{-1} \sum_{\mathbb{D}_j} |\K[\ell]|^{-2})^{-1/2}$. The estimated noise level, $\widehat \sigma = MAD(y_{J,k})/0.6745$ where $MAD$ is the median absolute deviations and $y_{J,k} = \innerp{y}{\Psi_{J,k}}$, the wavelet coefficients at the finest scale. The fine scale $j_1 = j_{1,1}$ is estimated using the default {\tt WaveD} method which is based on \cite{Cavalier-Raimondo-2007}.
	\item Compute the LRD {\tt WaveD} estimator which uses the same hard thresholding wavelet expansion. However, now the fine scale level and level dependent thresholds are modified. The fine scale level, $j_1 = j_{\alpha,1}$ is estimated using \eqref{eq:finescaleestimator} and the level dependent thresholds are estimated using $\widehat \lambda_j = \xi \tau_{\alpha,j} c_n$ where $c_n = \widehat \sigma \sqrt{\log n/n^{\alpha}}$ and $\tau_{\alpha,j}$ calculated using \eqref{eq:exactcov} and \eqref{eq:tau}. The smoothing parameter $\xi$ was tested for various different values. Similar to the i.i.d. case seen in \cite{Johnstone-et-al-2004}, the default bound given in the theory with $\xi \ge 2 \sqrt{2\alpha}$ was much too conservative. Instead the performance was found to be much better with the choice $\xi = \sqrt{\alpha}$. The presented results here are for the choices $\xi = \sqrt{\alpha}$ and $\xi = \sqrt{2\alpha}$, the latter being more effective with low levels of $\alpha$. In fact, the simulations were also conducted for $\xi = \sqrt{4\alpha}, \sqrt{6 \alpha}$ and $\sqrt{8 \alpha}$. These bigger smoothing parameters were only competitive for the cases when $\alpha \le 0.3$ and are omitted from the tables for brevity.
	\item Compute the empirical version of the MSE where,
	\[ 
		 \widehat{MSE}(\widehat f, f) = \widehat \Exp \norm{\widehat f - f}^2_2 = \frac{1}{M} \sum_{i = 1}^M \norm{\widehat f_i - f}_2^2.
	\]
\end{enumerate}
\begin{table}[ht]
\centering
\begin{tabular}{ccccccc}
\toprule
	\multicolumn{2}{c}{$DIP = 0.7$} & \multicolumn{5}{c}{$\alpha$} \\
	{\tt Signal} & {\tt Method}    & 1 & 0.8 & 0.6 & 0.4 & 0.2 \\
	\bottomrule 
Cusp & i.i.d. & 0.0056\hfill(3) & {\bf 0.0089}\hfill(3) & 0.0557\hfill(4) & 0.2525\hfill(4) & 0.6633\hfill(4) \\ 
10dB & $\xi = \sqrt{\alpha}$  & {\bf 0.0056}\hfill(3) & 0.0091\hfill(3) & 0.0238\hfill(3) & 0.0728\hfill(3) & 0.2235\hfill(3) \\ 
  & $\xi = \sqrt{2\alpha}$ & 0.0056\hfill(3) & 0.0091\hfill(3) & {\bf 0.0223}\hfill(3) & {\bf 0.0606}\hfill(3) & {\bf 0.1895}\hfill(3) \\
   \midrule
Cusp & i.i.d. & 0.0035\hfill(5) & 0.0039\hfill(5) & 0.0148\hfill(5) & 0.0650\hfill(5) & 0.1683\hfill(5) \\ 
20dB & $\xi = \sqrt{\alpha}$  & {\bf 0.0030}\hfill(5) & {\bf 0.0037}\hfill(4) & 0.0070\hfill(4) & 0.0099\hfill(3) & 0.0243\hfill(3) \\ 
  & $\xi = \sqrt{2\alpha}$ & 0.0039\hfill(5) & 0.0042\hfill(4) & {\bf 0.0054}\hfill(4) & {\bf 0.0084}\hfill(3) & {\bf 0.0206}\hfill(3) \\ 
   \midrule
Cusp & i.i.d. & 0.0013\hfill(6) & 0.0014\hfill(6) & 0.0036\hfill(6) & 0.0148\hfill(6) & 0.0392\hfill(6) \\ 
30dB & $\xi = \sqrt{\alpha}$  & {\bf 0.0011}\hfill(6) & {\bf 0.0014}\hfill(5) & 0.0025\hfill(5) & 0.0059\hfill(5) & 0.0069\hfill(4) \\ 
  & $\xi = \sqrt{2\alpha}$ & 0.0015\hfill(6) & 0.0016\hfill(5) & {\bf 0.0020}\hfill(5) & {\bf 0.0034}\hfill(5) & {\bf 0.0055}\hfill(4) \\
  \midrule
  LIDAR & i.i.d. & 0.0430\hfill(4) & 0.0418\hfill(4) & 0.0517\hfill(4) & 0.0972\hfill(4) & 0.4365\hfill(5) \\ 
10dB & $\xi = \sqrt{\alpha}$  & {\bf 0.0363}\hfill(4) & {\bf 0.0403}\hfill(4) & {\bf 0.0512}\hfill(3) & 0.0636\hfill(3) & 0.0993\hfill(3) \\ 
  & $\xi = \sqrt{2\alpha}$ & 0.0473\hfill(4) & 0.0488\hfill(4) & 0.0548\hfill(3) & {\bf 0.0633}\hfill(3) & {\bf 0.0913}\hfill(3) \\ 
   \midrule
  LIDAR & i.i.d. & 0.0128\hfill(5) & 0.0133\hfill(5) & {\bf 0.0167}\hfill(5) & 0.0483\hfill(6) & 0.1100\hfill(6) \\ 
20dB & $\xi = \sqrt{\alpha}$  & {\bf 0.0103}\hfill(5) & {\bf 0.0122}\hfill(5) & 0.0248\hfill(4) & {\bf 0.0299}\hfill(4) & 0.0393\hfill(4)\\ 
   & $\xi = \sqrt{2\alpha}$ & 0.0151\hfill(5) & 0.0164\hfill(5) & 0.0262\hfill(4) & 0.0305\hfill(4) & {\bf 0.0382}\hfill(4) \\ 
   \midrule
LIDAR & i.i.d. & 0.0049\hfill(6) & 0.0049\hfill(7) & 0.0060\hfill(7) & 0.0111\hfill(7) & 0.0230\hfill(7) \\ 
30dB & $\xi = \sqrt{\alpha}$  & {\bf 0.0041}\hfill(6) & {\bf 0.0044}\hfill(6) & {\bf 0.0057}\hfill(6) & 0.0088\hfill(6) & 0.0103\hfill(5) \\ 
   & $\xi = \sqrt{2\alpha}$ & 0.0053\hfill(6) & 0.0054\hfill(6) & 0.0059\hfill(6) & {\bf 0.0073}\hfill(6) & {\bf 0.0096}\hfill(5) \\ 
  \bottomrule
\end{tabular}
\caption{MSE for the Bumps and Cusp signals with $n = 4096$ and $M =1024$, the smoothing parameter for the LRD {\tt WaveD} method is $\xi = \sqrt{\alpha}$, the i.i.d. method uses the default $\eta = \sqrt{6}$. The typical estimated fine scale levels are shown in parentheses for each case.}
\label{table:MSE1}
\end{table}
  \begin{table}[ht]
  \centering
\begin{tabular}{ccccccc}
	\toprule
	\multicolumn{2}{c}{$DIP = 0.7$} & \multicolumn{5}{c}{$\alpha$} \\
	{\tt Signal} & {\tt Method}    & 1 & 0.8 & 0.6 & 0.4 & 0.2 \\
	\bottomrule
	Bumps & i.i.d.  & 0.7657\hfill(4) & 0.7717\hfill(4) & {\bf 0.7904}\hfill(4) & {\bf 0.8277}\hfill(4) & {\bf 0.9475}\hfill(5) \\ 
10dB & $\xi = \sqrt{\alpha}$  & {\bf 0.7615}\hfill(4) & {\bf 0.7705}\hfill(4) & 0.9376\hfill(3) & 0.9511\hfill(3) & 0.9831\hfill(3) \\ 
  & $\xi = \sqrt{2\alpha}$ & 0.7687\hfill(4) & 0.7768\hfill(4) & 0.9380\hfill(3) & 0.9517\hfill(3) & 0.9818\hfill(3) \\ 
   \midrule
Bumps & i.i.d.  & 0.5384\hfill(5) & {\bf 0.5405}\hfill(5) & {\bf 0.2905}\hfill(6) & {\bf 0.2702}\hfill(6) & {\bf 0.3165}\hfill(6) \\ 
20dB & $\xi = \sqrt{\alpha}$  & {\bf 0.5374}\hfill(5) & 0.5405\hfill(5) & 0.5469 \hfill(5) & 0.7583 \hfill(4) & 0.7661\hfill(4) \\ 
  & $\xi = \sqrt{2\alpha}$ & 0.5391\hfill(5) & 0.5417\hfill(5) & 0.5473\hfill(5) & 0.7583\hfill(4) & 0.7660\hfill(4) \\ 
   \midrule
Bumps & i.i.d. & 0.0793\hfill(7) & {\bf 0.0807}\hfill(7) & {\bf 0.0838}\hfill(7) & {\bf 0.0897}\hfill(7) & {\bf 0.0988}\hfill(7) \\ 
  30dB & $\xi = \sqrt{\alpha}$  & {\bf 0.0777}\hfill(7) & 0.2074\hfill(6) & 0.2093\hfill(6) & 0.21246\hfill(6) & 0.3141\hfill(6) \\ 
  & $\xi = \sqrt{2\alpha}$ & 0.0811\hfill(7) & 0.2080\hfill(6) & 0.2094\hfill(6) & 0.2118\hfill(6) & 0.3133\hfill(6) \\ 
  \midrule
Doppler & i.i.d.  & 0.0278\hfill(5) & {\bf 0.0293}\hfill(5) & {\bf 0.0369}\hfill(5) & {\bf 0.0631}\hfill(5) & 0.1199\hfill(5) \\ 
10dB & $\xi = \sqrt{\alpha}$  & {\bf 0.0263}\hfill(5) & 0.0472\hfill(4) & 0.0570\hfill(4) & 0.0689\hfill(4) & 0.1203\hfill(3) \\ 
  & $\xi = \sqrt{2\alpha}$ & 0.0292\hfill(5) & 0.0481\hfill(4) & 0.0568\hfill(4) & 0.0666\hfill(4) & {\bf 0.1196}\hfill(3) \\ 
   \midrule
Doppler & i.i.d. & 0.0103\hfill(6) & 0.0106\hfill(6) & {\bf 0.0122}\hfill(6) & {\bf 0.0183}\hfill(6) & 0.0455\hfill(7) \\ 
20dB & $\xi = \sqrt{\alpha}$  & {\bf 0.0096}\hfill(6) & {\bf 0.0104}\hfill(6) & 0.0200\hfill(5) & 0.0257\hfill(5) & 0.0304\hfill(5) \\ 
  & $\xi = \sqrt{2\alpha}$ & 0.0107\hfill(6) & 0.0110\hfill(6) & 0.0197\hfill(5) & 0.0248\hfill(5) & {\bf 0.0284}\hfill(5) \\ 
   \midrule
Doppler & i.i.d.  & 0.0029\hfill(7) & 0.0030\hfill(7) & {\bf 0.0032}\hfill(7) & 0.0043\hfill(8) & 0.0075\hfill(8) \\ 
30dB & $\xi = \sqrt{\alpha}$  & {\bf 0.0027}\hfill(7) & {\bf 0.0029}\hfill(7) & 0.0036\hfill(7) & 0.0055\hfill(7) & 0.0094\hfill(7) \\ 
  & $\xi = \sqrt{2\alpha}$ & 0.0031\hfill(7) & 0.0031\hfill(7) & 0.0033\hfill(7) & {\bf 0.0039}\hfill(7) & {\bf 0.0066}\hfill(7) \\ 
   \bottomrule
\end{tabular}
\caption{MSE for the Doppler and LIDAR signals with $n = 4096$ and $M =1024$, the smoothing parameter for the LRD {\tt WaveD} method is $\xi = \sqrt{\alpha}$, the i.i.d. method uses the default $\eta = \sqrt{6}$. The typical estimated fine scale levels are shown in parentheses for each case.}
\label{table:MSE2}
\end{table}
\subsection{Numerical results}
\label{ssec:numericalresults}
The results of the procedure are outlined in \autoref{table:MSE1} and \autoref{table:MSE2}. As stated earlier, the focus of the numerical study is the effect of the dependence structure so the DIP is fixed at $\nu = 0.7$. The most obvious fact is that overall the convergence rate tends to deteriorate as the level of dependence increases. This is consistent with the theoretical results in \autoref{sec:estimationmethod}.
 
The numerical results are not overall conclusive in favour of one method over the other for all noise levels. The main complication arises in the truncation of the wavelet expansion with the fine scale levels $\widehat j_1$ and $\widehat j_{\alpha,1}$. The typical fine scale levels (rounded to the nearest integer) are shown in \autoref{table:MSE1} and \autoref{table:MSE2} stated in parentheses for each case for each method. Due to the construction, $j_{\alpha,1} \le j_{1,1}$ for all $\alpha \in (0,1]$ and this is reflected in their estimates shown in \autoref{table:MSE1} and \autoref{table:MSE2}. As expected $\widehat j_{\alpha,1}$ decreases as $\alpha$ increases (stronger dependence) which is consistent with the theory. On the other hand, the na\"ive i.i.d. estimator increases as $\alpha$ decreases which can be either detrimental or beneficial to estimation as discussed below.

\begin{figure}[!ht]
\centering
\begin{minipage}{\columnwidth}
\subfloat[Doppler signal.]{
\label{comparisonfiga}
\includegraphics{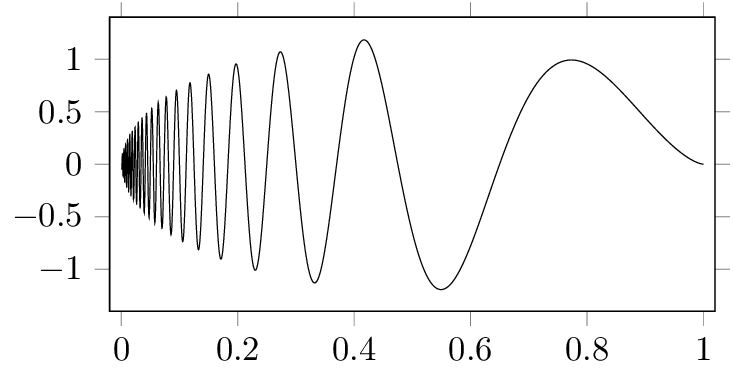}
}
\subfloat[Bumps signal.]{
\label{comparisonfigb}
\includegraphics{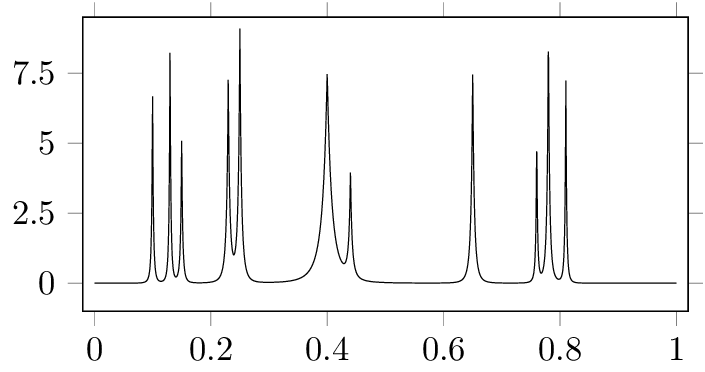}
}
\end{minipage}
\begin{minipage}{\columnwidth}
\subfloat[Blurred signal with $\nu = 0.7,\, \alpha = 0.5$.]{
\label{comparisonfigc}
\includegraphics{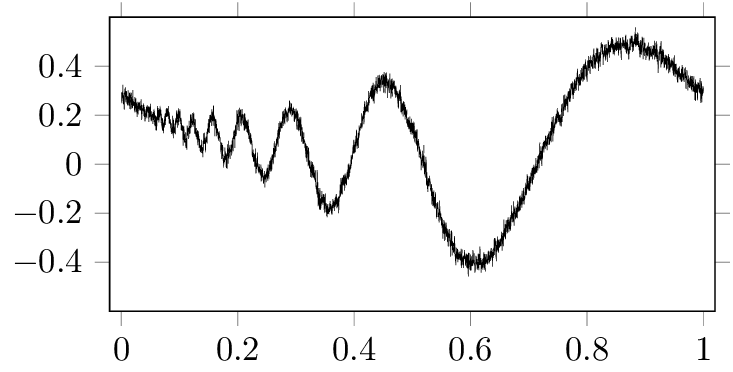}
}
\subfloat[Blurred signal with $\nu = 0.7,\, \alpha = 0.6$.]{
\label{comparisonfigc}
\includegraphics{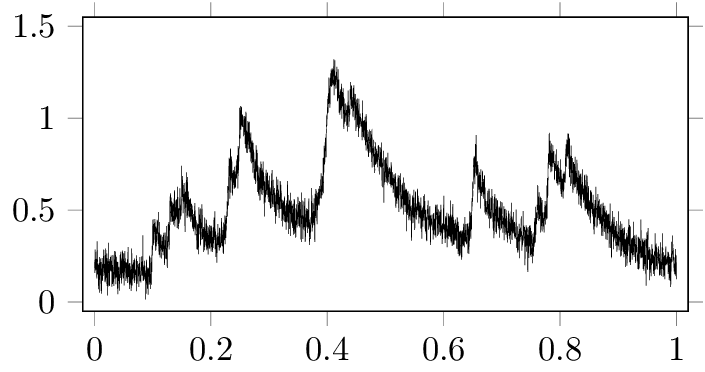}
}
\end{minipage}
\begin{minipage}{\columnwidth}
\subfloat[Default {\tt WaveD} reconstruction, $\widehat j_1 = 7$.]{
\label{comparisonfigc}
\includegraphics{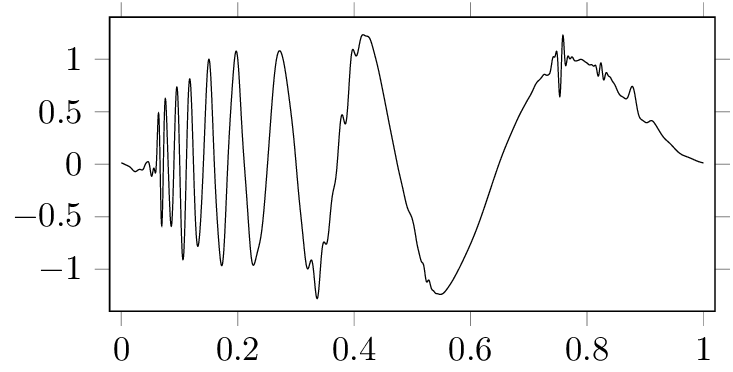}
}
\subfloat[Default {\tt WaveD} reconstruction, $\widehat j_1 = 5$.]{
\label{comparisonfigc}
\includegraphics{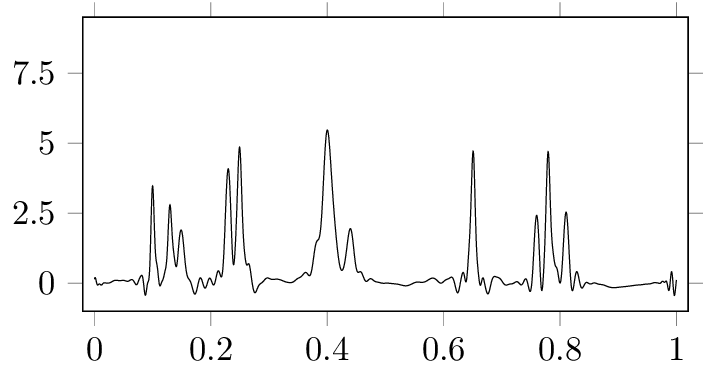}
}
\end{minipage}
\begin{minipage}{\columnwidth}
\subfloat[LRD {\tt WaveD} reconstruction, $\widehat j_{\alpha,1} = 6$.]{
\label{comparisonfigc}
\includegraphics{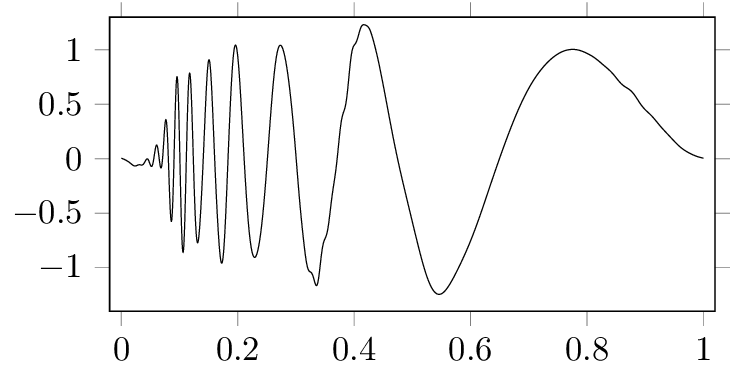}
}
\subfloat[LRD {\tt WaveD} reconstruction, $\widehat j_{\alpha,1} = 4$.]{
\label{comparisonfigc}
\includegraphics{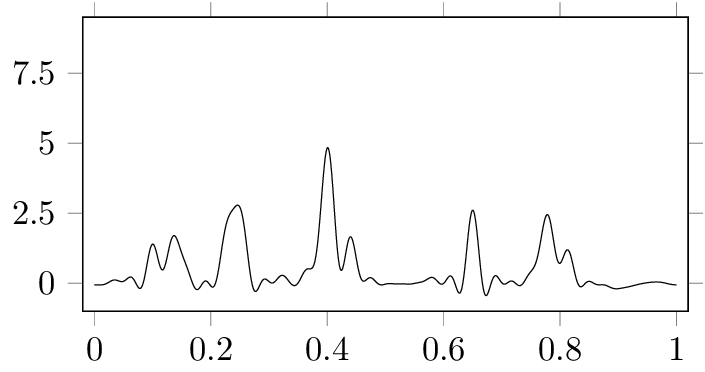}
}
\end{minipage}
\caption{Comparison of the LRD {\tt WaveD} method against the default {\tt WaveD} method. In both cases, $n = 4096 = 2^{12}$. The LRD method truncates the finest scale earlier as noted. For the Bumps and Doppler signals, SNR $\approx$ 20dB.}
\label{fig:2ndcomparison}
\end{figure}

In some cases the earlier truncation in the LRD method is favourable such as the LIDAR and Cusp signals with a strong level of dependence which is shown in \autoref{fig:comparisons}. The addition of higher scales to the wavelet expansion in the i.i.d. method does contribute to capturing more of the cusp feature and the last two peaks of the LIDAR signal. However, it is paid at the price that higher scales include spurious effects from the LRD noise, resulting in a poor estimator. This is reflected in \autoref{table:MSE1} where the MSE is smaller when $\widehat j_{\alpha,1} < \widehat j_1$ with the exception at the LIDAR signal at 20dB and $\alpha = 0.6$ and the Cusp signal at 10dB and $\alpha = 0.8$.

On the other hand, the earlier truncation of fine scale levels in the LRD method means that important features are not captured in the signal. This is seen in the Bumps and Doppler signals where the i.i.d. method tends to outperform the LRD method. In this situation the higher level scales in the expansion capture more features in the signal while not introducing the too much of the spurious LRD noise effects. A typical medium noise scenario demonstrating this is shown for the Bumps and Doppler signals in \autoref{fig:2ndcomparison}. Referring to the Doppler deconvolution in \autoref{fig:2ndcomparison} one can see that the spurious trends generated by the LRD noise are included in the default {\tt WaveD} reconstruction, however since $\widehat j_1 > \widehat j_{\alpha,1}$, the higher frequences of the Doppler signal can be captured. The LRD method does not include the spurious trends but pays the price of losing the higher frequencies in the Doppler signal. A similar behaviour is present in the Bumps signal. This behaviour is reflected in the numerical study in \autoref{table:MSE2} where the earlier truncation shows that the i.i.d. method outperforms the LRD method. However for the case of severely dependent noise at $\alpha = 0.2$ for the Doppler signal, the i.i.d. method loses its advantage and the addition of higher scales does not outperform the LRD method.

Therefore the LRD estimation method presented here offers an easily implementable solution that is resistant to the effects of long memory. The method is attractive to the case where the underlying function $f$ does not have a lot of transient high frequency behaviours where the higher scales of the wavelet transform are crucial. In the case where high frequencies are crucial to the signal, the established i.i.d. {\tt WaveD} method is perhaps more favourable since the signal features at higher scales are more important than the spurious noise.

\section{Proofs}
\label{sec:proof}
First the probabilistic result of the numerical estimation of the highest permissible scale level is given. Define the frequency levels,
	\begin{align}
		M_c &= \min\left\{ \ell, \ell > 0 : | \widetilde \K[\ell]| \le \ell^{\alpha/2} \varepsilon^{\alpha} \left(\log 1/\varepsilon^{2} \right)^{4/3} \right\}\label{eq:Mc}\\
		M_d &= \min\left\{ \ell, \ell > 0 : | \widetilde \K[\ell]| \le \ell^{\alpha/2} \varepsilon^{\alpha} \left( \log 1/\varepsilon^{2} \right)^{2/3}\right\}\label{eq:Md}.
	\end{align}

\begin{lemma}
	\label{lem:highlevel}
	Define the event, $\mathcal B = \cap_{\omega = 1}^{M_d} \left\{ |\varepsilon^\alpha \widetilde Z_H[\omega]| \le |\widetilde \K [\omega]|/2\right\}$. Then 
\[ 
	P(\mathcal{B}^c) \le c M_d \exp\left\{ - (\log 1/\varepsilon)^{1 + \xi}\right\}\eqqcolon \Omega_n
\] for some constants $c, \xi > 0.$ Further, define the event, $\mathcal M = \left\{ M_c \le M \le M_d\right\}$ then $P(\mathcal M^c) = \mathcal O (\Omega_n)$ and $M_c \le M_d \le M_1$ where $M_1 = \lfloor 2^{j_1}\rfloor $.
\end{lemma}

\textsc{Proof of \autoref{lem:highlevel}}. First we prove the statement that, $P(\mathcal B^c) \le \Omega_n$. By definition of $M_d$ in \eqref{eq:Md} there exists a $c \in (0,1)$ such that for all $\omega = 1,2,\ldots, M_d$; $| \widetilde \K[\omega]|/2 > \tfrac{c}{2}\ell^{\alpha/2} \varepsilon^{\alpha} \left( \log 1/\varepsilon \right)^{2/3}$. Thus, 
\begin{align*}
	P(\mathcal B^c) &= P\left({\displaystyle \bigcup_{\omega = 1}^{M_d}} \left\{ |\varepsilon^\alpha \widetilde Z_H[\omega]| > \frac{|\widetilde \K [\omega]|}{2}\right\} \right)\nonumber\\
	&\le \sum_{\omega= 1}^{M_d} P\left( |\varepsilon^\alpha \widetilde Z_H[\omega]| > |\widetilde \K [\omega]|/2\right) \nonumber\\
	&\le \sum_{\omega= 1}^{M_d} P\left( |\widetilde Z_H[\omega]| > \frac{c}{2}\omega^{\alpha/2} \left( \log 1/\varepsilon^{2} \right)^{2/3}\right) 
\end{align*}
From \eqref{eq:covZH}, $\Var \left(\widetilde Z_H[\omega] \right) \asymp |\omega|^{\alpha - 1}$ and apply the tail inequality for Gaussian random variables, let $Z \sim \mathcal N(0,1)$, then there exists a constant $C > 0$ such that,
\begin{align}
	P(\mathcal B^c) &\le \sum_{\omega= 1}^{M_d} P\left( |Z| > c \omega^{1/2} \left( \log 1/\varepsilon^{2} \right)^{2/3}\right) \nonumber\\
	&\le C M_d\exp \left\{ - \frac{1}{2}c^2  \left( \log 1/\varepsilon\right)^{4/3}\right\} = \mathcal O(\Omega_n).\label{eq:PBc}
\end{align}
For $\mathcal M^c$ the event can be written $\mathcal M^c = \left\{ M > M_d \right\} \cup \left\{ M < M_c \right\}$ and start by considering the first scenario,
\begin{align}
	P(M > M_d) &\le P(\cap_{\omega = 1}^{M_d} \left\{ |\widetilde Y_e[\omega]| >  \omega^{\alpha/2} \varepsilon^{\alpha} \log 1/\varepsilon^2 \right\})\nonumber\\
		&\le P(|\widetilde Y_e[M_d]| >  M_d^{\alpha/2} \varepsilon^{\alpha} \log 1/\varepsilon^2 )\nonumber\\
		&\le P(|\widetilde Y_e[M_d]| >  M_d^{\alpha/2} \varepsilon^{\alpha} \log 1/\varepsilon^2, \, \mathcal B ) + P(\mathcal B^c)\label{eq:ineq1}
\end{align}
Now under the event $\mathcal B$ with \eqref{eq:regular}, $|\widetilde Y_e[M_d] | \le \frac{3}{2}|\widetilde K[M_d]| \le C|M_d|^{-\nu}$. Further, by definition of $M_d$ and \eqref{eq:regular}
\begin{equation}
	M_d \asymp \left(\varepsilon^{\alpha} \left(\log1/\varepsilon\right)^{2/3} \right)^{-2/(2\nu+ \alpha)}\label{eq:highscale1o}
\end{equation}
which means that as $n \to \infty$, $M_d^{-\nu -\alpha/2} = o(\varepsilon^{\alpha} (\log 1/\varepsilon^2) )$. Hence, from \eqref{eq:ineq1} with \eqref{eq:highscale1o} and \eqref{eq:PBc}, there is an $N \in \mathbb{Z}^+$ such that for all $n \ge N$,
\begin{equation}
	P(M > M_d) \le  P(M_d^{-\nu - \alpha/2} >  \varepsilon^{\alpha} \log 1/\varepsilon^2) + P(\mathcal B^c) = P(\mathcal B^c) = \mathcal O(\Omega_n).\label{eq:M1}
\end{equation}
On the other hand,
\begin{align*}
	P(M < M_c) &\le P\left(\cup_{\omega = 1}^{M_c-1} \left\{ |\widetilde Y_e[\omega]| \le \omega^{\alpha/2} \varepsilon^{\alpha} \log 1/\varepsilon^2 \right\} \right) \nonumber\\
	&\le \sum_{\omega = 1}^{M_c-1} P\left( |\widetilde Y_e[\omega]| \le \omega^{\alpha/2} \varepsilon^{\alpha} \log 1/\varepsilon^2  \right).
\end{align*}
Under the event $\mathcal B$, $|\widetilde Y_e[\omega]| \ge |\widetilde K[\omega]|/2 \ge c M_c^{-\nu}$ for all $\omega = 1,2,\ldots, M_c-1.$ Similar to \eqref{eq:highscale1o}, by \eqref{eq:Mc} and \eqref{eq:regular},
\begin{equation}
	M_c \asymp \left( \varepsilon^{\alpha} \left(\log 1/\varepsilon^2\right)^{4/3} \right)^{ - 2/(2 \nu + \alpha)}\label{eq:highscale2o}
\end{equation}
and consequently, as $n \to \infty$, $\varepsilon^{\alpha} (\log1/\varepsilon^2) = o(M_c^{-\nu - \alpha/2})$. Hence from \eqref{eq:PBc}, 
\begin{equation}
	P(M < M_c) \le P(\mathcal B^c) = \mathcal O(\Omega_n).\label{eq:M2}
\end{equation}
Therefore, \eqref{eq:M1} with \eqref{eq:M2} yields the first result of the Lemma. The second result of the Lemma follows from \eqref{eq:highscale1o} and \eqref{eq:highscale2o}.\qed

\subsection{Maxiset Theorem}
\label{sec:Maxi}
As in similar previous works in \cite{Kulik-Raimondo-2009a,Johnstone-et-al-2004} the proof of the main result imitates the same maxiset approach. Roughly speaking, this approach finds the `maxiset' class of functions for a general hard thresholding wavelet estimator. This Maxiset theorem is stated here for easy reference.  First, introduce the notation: $\mu$ will denote the measure such that for $j \in \mathbb{N}$, $k \in \mathbb{N}$,
\begin{align*}
		\mu\left\{ (j,k)\right\} &= \norm{\sigma_{j,\nu,\alpha}\Psi_{j,k}}_p^p = \sigma_{j,\nu,\alpha}^p 2^{j(\tfrac{p}{2}-1)}\norm{\Psi}_p^p,\\
		l_{q,\infty}(m) &= \left\{ f, \sup_{\lambda > 0} \lambda^q \mu\left\{ (j,k): \left| \beta_{j,k} \right| > \sigma_{j,\nu,\alpha} \lambda \right\} < \infty \right\}.
\end{align*}
\begin{theorem}[Maxiset]
\label{Maxiset}
Let $p > 0$, $0 < q < p$ and $\left\{ \Psi_{j,k}, j \ge -1, k = 0,1,\ldots, 2^{j-1}\right\}$ be a periodised wavelet basis of $\mathscr L_2([0,1])$ and $\sigma_j$ is a positive sequence such that the heteroscedastic basis $\sigma_j \Psi_{j,k}$ satisfies the Temlyakov property. Suppose that the index set $\Lambda_n$ is a set of pairs ($j,k$) and $c_n$ is a deterministic sequence tending to zero with,
\begin{equation}
	\sup_n \mu(\Lambda_n ) c_n^p < \infty.\label{eq:Resolutiontuning}
\end{equation}
If for any $n$ and any pair $(j,k) \in \Lambda_n$ we have,
\begin{equation}
	\Exp|\widehat \beta_{j,k} - \beta_{j,k}|^{2p} \le C(\sigma_j c_n)^{2p};\qquad P(|\widehat \beta_{j,k} - \beta_{j,k}| \ge \xi \sigma_j c_n/2) \le C(c_n^{2p} \wedge c_n^4) \label{eq:coefconditons}
\end{equation}
for some positive constants $\xi$ amd $C$ then the wavelet estimator,
\[
	\widehat f_n(x) = \sum_{(j,k) \in \Lambda_n} \widehat \beta_{j,k} \Psi_{j,k}(x) \mathbbm{1}_{\left\{ |\widehat \beta_{j,k}| \ge \xi \sigma_j c_n\right\}}
\]
satisfies the following for all positive integers $n$,
\[
	\Exp \norm{\widehat f_n - f}_p^p \le C c_n^{p-q}.
\]
if and only if,
\begin{equation}
	f \in l_{q,\infty}(\mu) \quad \text{and} \quad \sup_{n\ge1}\left( c_n^{q-p} \norm{ f - \sum_{j,k \in \Lambda_n} \beta_{j,k}\Psi_{j,k}}_p^p\right) < \infty.\label{eq:maxicondition}
\end{equation}

\end{theorem}
The proof of the rate result in \autoref{Theorem} will use the Maxiset Theorem by verifying the regularity conditions, \eqref{eq:Resolutiontuning}, \eqref{eq:coefconditons} and \eqref{eq:maxicondition}, then choosing $q$ for the {\it dense} and {\it sparse} regions respectively with,
\begin{alignat}{2}
	q &= q_d \coloneqq \frac{p(2\nu + \alpha)}{2s + 2\nu + \alpha} &&\quad \text{when} \quad s \ge (\nu + \alpha/2)(p/\pi - 1)\label{eq:qdense}\\
	q &= q_s \coloneqq \frac{(2\nu + \alpha)p - 2}{2s + 2\nu + \alpha - \frac{2}{\pi}}  &&\quad  \text{when }1/\pi - \alpha/2 - \nu \le s < ( \nu + \alpha/2)(p/\pi- 1)\label{eq:qsparse}
\end{alignat}

\subsection{Stochastic analysis of the estimated wavelet coefficients}
\label{ssec:waveletbounds}

By definition, it is clear that the estimated wavelet coefficients have no bias.  Consider now the covariance structure of the $\widetilde{Z_H}$ process,
\[
	\Cov \left( \widetilde{ Z_H}[\omega], \widetilde{Z_H}[\ell] \right) = \Exp  \widetilde{ Z_H}[\omega] \overline{\widetilde{Z_H}}[\ell] =  \Exp \int_\mathbb{R}  e^{- 2 \pi i \omega x}  \,dB_H(x)\, \int_\mathbb{R} e^{2 \pi i \ell y} \,dB_H(y).
\]
To evaluate this, appeal to Theorem 2 of \cite{Wang-1996} which uses a representation of the fractional Gaussian noise process via a Wavelet-Vaguelette Decomposition (WVD)
\[
	dB_H(x) = \sum_{j,k} \xi_{j,k} u_{j,k}(x)\, dx,
\]
where $\xi_{j,k}$ is a white noise process and $u_{j,k}$ is a set of vaguelette basis functions defined with,
\[
	u_{j,k}(x) \coloneqq (-\Delta)^{1/4-H/2} \varphi_{j,k}(x),
\]
where $\Delta = \frac{d^2}{dx^2}$ is an elliptic operator and $\varphi_{j,k} \map{\mathbb{R}}{\mathbb{R}}$ is a set of orthogonal wavelet functions. Using this representation of fractional Gaussian noise we can write,
\begin{align}
	\Cov \left( \widetilde{ Z_H}[\omega], \widetilde{Z_H}[\ell] \right) &=  \sum_{j,k} \sum_{j',k'}  \Exp \xi_{j,k} \xi_{j',k'} \int_\mathbb{R}  e^{- 2 \pi i \omega x}  u_{j,k}(x)\, dx\, \int_\mathbb{R} e^{2 \pi i \ell y} u_{j',k'}(y)\, dy.\nonumber\\
	&= \sum_{j,k} \widetilde{u_{j,k}}[\omega]\overline{\widetilde{u_{j,k}}[\ell]}.\label{eq:CovZ1}
\end{align}
The operator $(-\Delta)^{-(H-1/2)/2}$ for $H \in (1/2,1)$ is known from the theory of singular integrals as the \textit{Reisz Potential} \citep[see for example][Chapter V]{Stein-1970} and has the representation,
\begin{equation}
	 (-\Delta)^{-\tfrac{H-1/2}{2}} f(x) = \frac{\Gamma\left(\tfrac{3}{4}- \tfrac{H}{2}\right) }{\sqrt{\pi} 2^{H-\tfrac{1}{2}}\Gamma \left( \tfrac{H}{2} - \tfrac{1}{4}\right)}\int_\mathbb{R} f(z) |z-x|^{H - 3/2}\, dz.\label{eq:RieszPotential}
\end{equation}
For our purposes its behaviour in the Fourier domain has been evaluated by \cite*{Samko-Kilbas-Marichev-1993}. Indeed, apply Theorem 12.2 of \cite{Samko-Kilbas-Marichev-1993} with \eqref{eq:RieszPotential} then for any $f \in \mathscr L_1(\mathbb{R})$,
\begin{align}
	\mathcal F (-\Delta)^{1/4-H/2} f(\omega) &= \widetilde f(\omega)| \omega|^{1/2 - H}\label{eq:conv}.
\end{align}
From \eqref{eq:CovZ1} and \eqref{eq:conv}, it would be desirable to use a wavelet function $\varphi \in \mathscr L_1(\mathbb{R})$ that has a simple behaviour in the Fourier domain. Naturally, a suitable choice is the Meyer wavelet, $\varphi = \psi$, since it is bandlimited (see \autoref{ssec:Meyer}) and makes the calculations easier. Therefore we have,
\begin{equation}
	\Cov \left( \widetilde{Z_H}[\omega], \widetilde{Z_H}[\ell]\right) = |\omega \ell|^{1/2 - H} \sum_{j,k} \widetilde{\psi_{j,k}}(\omega) \overline{\widetilde{\psi_{j,k}}(\ell)}.\label{eq:exactcov}
\end{equation}
Consider an arbitrary $\omega \in \mathbb{Z}$ and consider the possible values of $j$ such that $\omega \in \mathbb{D}_j$. The summation sets will have a non empty intersection, $\mathbb{D}_j \cap \mathbb{D}_{j'} \neq \emptyset$, if and only if $j' \in \left\{ j-1,j,j+1 \right\}$. Therefore the summands in \eqref{eq:exactcov} will be nonzero for, at most, three different values of $j$. The Meyer wavelet also is bounded with, $|\widetilde \psi| \le 1$. Therefore we can crudely bound the magnitude of the covariance with,
\begin{align}
	\left|\Cov \left( \widetilde{Z_H}[\omega], \widetilde{Z_H}[\ell]\right)\right| &=  | \omega\ell|^{1/2 - H} \left|\sum_{j \in \mathbb{Z}} \sum_{k = 0}^{2^j-1} 2^{-j} e^{2^{1-j}\pi i (\ell-\omega) k}\widetilde{\psi}(\omega 2^{-j}) \overline{\widetilde{\psi}(\ell 2^{-j})}\right|\nonumber\\
	&\le 3   |\omega \ell|^{1/2 - H}.\label{eq:covZH} 
\end{align}
Thus we are in a position to bound the variance of the estimated wavelet coefficients,
\begin{align}
	\Var \left(\widehat \beta_{j,k} \right) &= \varepsilon^{2\alpha}  \Var \left(\sum_{\ell \in \mathbb{D}_j}  \frac{\widetilde{ Z_H}[\ell]}{\widetilde \K[\ell]} \overline{\widetilde{\Psi_{j,k}}}[\ell]\right)\nonumber\\
		&=  \varepsilon^{2\alpha} \sum_{\omega \in \mathbb{D}_j} \sum_{\ell \in \mathbb{D}_j} \frac{\overline{\widetilde{\Psi_{j,k}}}[\ell] \widetilde{\Psi_{j,k}}[\omega]}{\widetilde \K[\ell] \overline{\widetilde \K}[\omega]}\Cov \left(\widetilde{Z_H}[\omega],\widetilde{Z_H}[\ell] \right) \eqqcolon \varepsilon^{2\alpha}\tau_{\alpha,j,k}^2. \label{eq:tau}
		\end{align}
To gain insight to the overall asymptotic structure, the variance of the coefficients needs to be bounded. Use \eqref{eq:covZH} and that the cardinality of $\mathbb{D}_j$, $|\mathbb{D}_j| = 2^{j+1}$,
\begin{align*}
		\tau^2_{\alpha,j} &\le3    \sum_{\omega \in \mathbb{D}_j}  \left| \frac{ \widetilde{\Psi_{j,k}}[\omega]}{ \overline{\widetilde \K}[\omega]} \right| |\omega|^{1/2-H} \sum_{\ell \in \mathbb{D}_j} \left| \frac{\overline{\widetilde{\Psi_{j,k}}}[\ell] }{\widetilde \K[\ell] } \right|| \ell|^{1/2 - H} \nonumber\\
		&= 3    \left( \sum_{\omega \in \mathbb{D}_j}  \left| \frac{ \widetilde{\Psi_{j,k}}[\omega]}{ \overline{\widetilde \K}[\omega]} |\omega|^{1/2-H}\right| \right)^2  \nonumber\\
		&\le 3    \left| \mathbb{D}_j \right| 2^{-j} \sum_{\omega \in \mathbb{D}_j}  \left| \frac{ \widetilde{\Psi}[\omega 2^{-j}]}{ \widetilde \K[\omega]}  \right|^2 |\omega|^{1-2H}     \nonumber\\
		&\le 6     \sum_{\omega \in \mathbb{D}_j}  \left| \frac{ \widetilde{\Psi}[\omega 2^{-j}]}{ \widetilde \K[\omega]}  \right|^2 |\omega|^{1-2H}\nonumber\\
		&\le 6  \sup_{x \in \mathbb{D}_j} |x|^{1-2H} \sup_{y \in \mathbb{D}_j} |\widetilde K[y]|^{-2} \int_\mathbb{R} |\widetilde \Psi(z)|^2 \, dz \nonumber\\
		&\le C  2^{-j(1-\alpha-2\nu)} \norm{\Psi}_2^2,
\end{align*}
where the last two lines follow by Parseval and Plancherels identities and condition \eqref{eq:regular}. Thus, the asymptotic behaviour of the variance of the wavelet coefficients at scale $j$ are bounded with,
\[ \Var \left(\widehat \beta_{j,k} \right) = \varepsilon^{2\alpha}\tau_{\alpha,j}^2 \le C\varepsilon^{2\alpha} 2^{-j(1-\alpha-2\nu)} \le C \sigma_{j,\nu,\alpha}^2 c_n^2,\]
where $\sigma_{j,\nu,\alpha}$ and $c_n$ are defined in \eqref{eq:scalefactor} and \eqref{eq:samplefactor}. Since $\widehat \beta_{j,k}$ is Gaussian, then from the variance bound it follows that, 
\[\Exp \left| \widehat{\beta}_{j,k} - \beta_{j,k}\right|^{2p} \le C_p (\sigma_{j,\nu,\alpha} c_n)^{2p} . \]

Let $\xi \ge \sqrt{4\alpha (p \vee 2)}$ and $Z \sim \mathcal N(0,1)$, then from the tail inequality for Gaussian random variables. 
\begin{align*}
	P\left( \left| \widehat{\beta}_{j,k} - \beta_{j,k}\right| \ge \frac{\xi \tau_{j,\nu,\alpha} c_n}{2} \right) &= 2P\left(  Z  \ge \frac{\xi \sqrt{\log n}}{2} \right)\\
		&\le \frac{2\sqrt{2}}{\xi \sqrt{ \pi \, \log n} } \exp\left\{ - \frac{\xi^2 \log n}{4} \right\}\\
		&\le C_{\xi} n^{ - \tfrac{\xi^2}{4}} \le C (c_n^{2p} \wedge c_n^4).
\end{align*}
This verifies the wavelet coefficient conditions in \eqref{eq:coefconditons}.

\subsection{Temlyakov Property and resolution tuning}
\label{ssec:Temlyakov}

Recall $\Lambda_n = \left\{ (j,k)  : -1 \le j \le j_{\alpha,1}; k = 0,1,\ldots, 2^j-1 \right\}$. Consider the set of scales  $\Lambda_n^1 = \left\{ j \in \mathbb{Z}: -1 \le j \le j_{\alpha,1}, j_{\alpha,1} \in \mathbb{Z}^+ \right\}$. Then from \citet[Appendices A.1 \& B.2]{Johnstone-et-al-2004}, the heteroskedastic basis $\left\{ \sigma_j, \Psi_{j,k}(\cdot)\right\}$ satisfies the Temlyakov property as soon as,
\[ \sum_{j \in \Lambda_n^1}2^j \sigma_j^2 \le C \sup_{j \in \Lambda_n^1 } \left( 2^j \sigma_j^2 \right) \qquad \text{and} \qquad \sum_{j \in \Lambda_n^1}2^{jp/2} \sigma_j^2 \le C \sup_{j \in \Lambda_n^1 } \left( 2^{jp/2} \sigma_j^p \right).\]
This property is satisfied in our framework with $\sigma_j = \sigma_{j,\nu,\alpha} \le C 2^{-j/2(1-\alpha-2\nu)}$. Now verify the resolution tuning condition \eqref{eq:Resolutiontuning}.
\begin{align*}
	\mu(\Lambda_n) &= \sum_{j \le j_{\alpha,1}} 2^j \sigma_{j,\nu,\alpha}^p 2^{j(p/2-1)}\norm{\Psi}_p^p < C  2^{j_{\alpha,1} p(\alpha+2\nu)/2}.
\end{align*}
Using the choice, $2^{j_{\alpha,1}} \asymp \left( n^\alpha/\log n \right)^{1/(\alpha+2\nu)}$ yields $\mu(\Lambda_n) c_n^p = \mathcal O(1)$. Consequently, \eqref{eq:Resolutiontuning} is verified.

\subsection{Besov Embedding}
\label{ssec:Besov}
One needs to find a Besov scale such that $\Besov_{\pi, r}^\delta \subseteq l_{q,\infty}$ and that the maxiset condition \eqref{eq:maxicondition} holds. As shown in \cite{Johnstone-et-al-2004}, the problem can be simplified by considering the the set of functions $l_q(\mu) \subset l_{q,\infty}(\mu)$ where,
\[ l_q(\mu) \coloneqq \left\{ f \in \mathscr L_p : \sum_{j,k \in A_j} \frac{\left|\beta_{j,k}\right|^q}{\sigma_j^q}\norm{\sigma_j \Psi_{j,k}}_p^p < \infty \right\}, \]
where $A_j$ is a set of cardinality proportional to $2^j$. Since $\norm{\sigma_j \Psi_{j,k}}_p^p = \sigma_j^p 2^{j(\tfrac{p}{2}-1)}$, then $f \in l_q(\mu)$ if
\[ 
	\sum_{j \ge 0} 2^{\tfrac{j}{2}\left( p -2 -(p-q)(1-\alpha - 2\nu )\right)}\sum_{k = 0}^{2^j -1}|\beta_{j,k}|^q  =  \sum_{j \ge 0} 2^{jq\left( (\nu + \tfrac{\alpha}{2})(\tfrac{p}{q}-1) + \tfrac{1}{2} - \tfrac{1}{q}\right)}\sum_{k = 0}^{2^j -1}|\beta_{j,k}|^q < \infty. 
\]
This condition is true for $f \in \Besov_{q,q}^\delta$ with the choice $\delta = (\nu + \tfrac{\alpha}{2})(\tfrac{p}{q}-1). $
Then depending on whether we are in the dense or sparse case the levels $s, \pi$ and $r$ are determined such that $\Besov^s_{\pi,r} \subseteq \Besov_{q,q}^\delta$. There exists two Besov embeddings given by,
\begin{align}
	\mathcal B^s_{\pi, r} &\subseteq \mathcal B^{\gamma}_{\rho, r} \qquad \text{when}\quad 0 <  \rho \le \pi  \text{ and }s \ge \gamma; \label{eq:embedA}\\
	\mathcal B^s_{\pi, r} &\subseteq \mathcal B^{\gamma}_{\rho, r} \qquad \text{when}\quad \pi < \rho \text{ and }s - 1/\pi =  \gamma - 1/\rho. \label{eq:embedB}
\end{align}

\textbf{The dense phase.} Choose the level of $q = q_d$ where,
\[
	q_d \coloneqq \frac{p(2\nu + \alpha)}{2s + 2\nu + \alpha} \qquad \text{when} \quad s \ge (\nu + \alpha/2)(p/\pi - 1).
\]
Then find the levels $s, \pi$ and $r$ such that $\Besov^s_{\pi,r} \subseteq \Besov_{q,q}^\delta$ where $s \ge (\nu + \alpha/2 ) (p/\pi-1)$.

By definition, one has, $s \ge \left(\alpha/2+ \nu \right)\left( p/\pi - 1\right)$. Eliminate $p$ by substituting $p = (2s + 2\nu + \alpha)q_d/(2\nu + \alpha)$ yields, $\pi \ge q_d$. So we need to prove $s \ge \delta =  \left(2\nu + \alpha\right)\left(p/2q_d- 1/2\right)$ to be able to use \eqref{eq:embedA}.  However, by definition of $q_d$ we have, $\delta = \left(\nu + \alpha/2\right)\left(p/q_d- 1\right) = s > 0$ by assumption.

 \textbf{The sparse phase.} Choose the level of $q = q_s$ where,
\[
q_s \coloneqq \frac{(2\nu + \alpha)p - 2}{2s + 2\nu + \alpha - \frac{2}{\pi}}  \qquad \text{when} \quad 1/\pi - \alpha/2 - \nu < s < ( \nu + \alpha/2)(p/\pi- 1).
\]
To ensure the inequalities in the above equation are valid, it requires that $p > 2/(2\nu + \alpha)$. Then $\delta =  \left(2\nu + \alpha\right)\left(p/2q_s- 1/2\right)$ and we have,
\begin{align*}
	\delta &= \left(2\nu + \alpha\right)\left(p/2q_s- 1/2\right)\\
	&= \frac{(2 \nu + \alpha)(sp-p/\pi + 1)}{(2\nu + \alpha)p-2}.
\end{align*}
Consider the scenario when $\pi \ge q_s$ and use embedding \eqref{eq:embedA}. This requires that $s \ge \delta  = (2 \nu + \alpha)(sp-p/\pi + 1)/((2\nu + \alpha)p-2)$, or equivalently, $s \le (\nu + \alpha/2)(p/\pi - 1).$ It is also needed that $\delta > 0 $, which implies that either $(i)$ $p > 2/(2\nu + \alpha)$ and $s > 1/\pi - 1/p$; or $(ii)$ $p < 2/(2\nu + \alpha)$ and $s < 1/\pi - 1/p$. The $(ii)$ scenario is impossible since it is assumed that $p \ge 1$ and $s \ge 1/\pi$  which contradicts $s < 1/\pi - 1/p$. The condition $p > 2/(2\nu + \alpha)$ is verified since by assumption in the sparse phase, $\nu \ge 1 - \alpha/2$ and $p > 1$. Then $(i)$ implies that, $s \ge 1/\pi - \nu - \alpha/2$. 

Consider now the scenario when $\pi < q_s$ by definining,
\begin{equation}
	s' = s - 1/\pi + 1/q.
	\label{eq:sdash}
\end{equation}
Then use the embedding \eqref{eq:embedB}. Indeed, if we solve \eqref{eq:sdash} with $s' = \delta$, then $q = q_s$ and the embedding of \eqref{eq:embedB} applies.
 
To apply \autoref{Maxiset}, \eqref{eq:maxicondition} needs to be verified. Therefore we need to find a $\delta > 0$ such that for any $f \in \Besov_{p,r,}^\delta$, \eqref{eq:maxicondition} is satisfied.
\begin{align*}
	c_n^{q-p} \norm{f - \sum_{j,k }\beta_{j,k}\Psi_{j,k}}_p^p = c_n^{q-p}2^{-j_{\alpha,1} \delta p} \norm{f}_{\Besov_{p,r}^\delta} = c_n^{q-p+ 2\delta p/(2\nu + \alpha)} \norm{f}_{\Besov_{p,r}^\delta}.
\end{align*}
The above is bounded uniformly in $n$ if we choose $\delta = 1/2(2\nu + \alpha)(1 - q/p)$. Now we need to find $s,\pi$ such that $\Besov_{\pi,r}^s\subseteq \Besov_{p,r}^\delta$.

Consider the first case $\pi \ge p$. This case cannot occur in the sparse phase due to \eqref{eq:sparse} and the assumption that $s$ is positive. In the dense phase, use embedding \eqref{eq:embedA} with $\gamma = \delta$ and $q = q_d$. Therefore, \eqref{eq:embedA} holds if $s \ge 1/2(2\nu + \alpha)(1 - q_d/p)$. This implies,
\begin{align*}
	s &\ge 1/2(2\nu + \alpha)(1 - q_d/p)\\
	&= \frac{(2\nu + \alpha)s}{2s + 2\nu + \alpha},
\end{align*}
which always holds under the assumption that $s > 0$.

Now consider the dense case when $\pi < p$. In this scenario use embedding \eqref{eq:embedB} by defining $s - 1/\pi = \gamma - 1/p$ which ensures $\Besov_{\pi,r}^s \subseteq \Besov_{p,r}^\gamma$. Then complete the embedding using \eqref{eq:embedA} (namely, $\Besov_{p,r}^\gamma \subseteq \Besov_{p,r}^\delta$) which requires $\gamma \ge \delta$ with $q = q_d$ or equivalently after rearrangement, $2s\gamma + (2\nu + \alpha)(1/p - 1/\pi) \ge 0.$ The left hand side is greater than $(s - 1/\pi)(p/\pi - 1)(2\nu + \alpha) \ge 0$ when $s \ge (\nu + \alpha/2)(p/\pi -1)$ (which is true in the dense phase).

The last case to consider is the sparse case when $\pi < p$. Again introduce a new Besov scale $\gamma$ defined with, $s - 1/\pi = \gamma - 1/p$ and apply a similar argument to above which requires that, $\gamma \ge \delta$ with $q = q_s$. This is satisfied if $s > 1/\pi$, which always holds.

\subsection{Proof of Theorem 1}
\label{ssec:ProofThm1}

The proof of the theorem is an application of \autoref{Maxiset} with the choice of $\sigma_j = \sigma_{j,\nu,\alpha}$ and $c_n$ and $\xi$ defined in \autoref{Maxiapproach} and the arguments in \autoref{ssec:waveletbounds}, \autoref{ssec:Temlyakov} and \autoref{ssec:Besov}. The dense regime rate result of \eqref{eq:dense} is derived with the choice of $q = q_d$ in \eqref{eq:qdense} and the Besov embedding argument in \autoref{ssec:Besov}. Similarly, the sparse rate regime in \eqref{eq:sparse} is derived with the Besov embedding result in \autoref{ssec:Besov} with the choice of $q = q_s$ in \eqref{eq:qsparse}.

\section*{Acknowledgements}
I would like to acknowledge The University of Sydney since part of this work was achieved there. I would also like to thank Rafa{\l} Kulik for helpful comments. This research was partially supported under Australian Research Council's Discovery Projects funding scheme (project number DP110100670). 

\def\cprime{$'$}

\end{document}